# The JEM-EUSO Mission

Toshikazu Ebisuzaki, H. Mase, Y. Takizawa, Y. Kawasaki, and K. Shinozaki

*RIKEN Advanced Science Institute, 2-1 Hirosawa, Wako351-0198, Japan*

F. Kajino

*Department of Physics, Konan University, Okamoto 8-9-1, Higashinada, Kobe 658-8501, Japan*

N. Inoue

*Graduate School of Science and Engineering, 255 Shimo-Okubo, Sakura-ku, Saitama City, Saitama 338-8570, Japan*

N. Sakaki

*College of Science and Engineering, Aoyama Gakuin University, 5-10-1 Fuchinobe, Chuo-ku, Sagamihara-shi, Kanagawa 229-8558, Japan*

A. Santangelo

*Astronomie und Astrophysik, Universitt Tubingen, Sand 1, 72076 Tubingen, Germany*

M. Teshima,

*Max-Planck-Institut for Physik, Fohringer Ring 6, 80805 Munich, Germany*

E. Parizot and P. Gorodetzky,

*APC, Univ. of Paris Diderot, CNRS/IN2P3, 10, rue A. Domon et L. Duquet, 75205 Paris Cedex 13, France*

O. Catalano

*Istituto di Astrofisica Spaziale e Fisica Cosmica di Palermo, INAF, Via Ugo La Malfa 153, 90146 Palermo, Italy*

P. Picozza and M. Casolino,

*Department of Physics, University of Rome Tor Vergata, Via della Ricerca Scientifica 1, 00133 Rome, Italy*

M. Panasyuk and B.A. Khrenov

*SINP, Lomonosov Moscow State Univ., Leninskie Gory 1 str. 2, Moscow, 119991, Russia*

I.H. Park

*Department of Physics, Ewha Womans University, Seoul 120-750, Korea*

T. Peter

*Institute for Atmospheric and Climate Science, ETH Zurich, Zurich, Switzerland*

G. Medina-Tanco

*Inst. de Ciencias Nucleares, UNAM, AP 70-543 / CP 04510, Mexico D.F.*

D. Rodriguez-Frias

*University of Alcala Ctra. Madrid-Barcelona, km. 33.6, E-28871, Alcala de Henares, Madrid, Spain*

J. Szabelski,

*Soltan Institute for Nuclear Studies, 90-950 Lodz, Box 447, Poland*

P. Bobik

*Institute of Experimental Physics SAS, Watsonova 47, 040 01 Kosice, Slovakia*

(for the JEM-EUSO collaboration)

The JEM-EUSO mission explores the origin of the extreme energy cosmic rays (EECRs) above 100 EeV and explores the limits of the fundamental physics through the observations of their arrival directions and energies. It is designed to achieve an exposure larger than 1 million $km^2$ sr year to open a new particle astronomy channel. This super-wide-field (60 degrees) telescope with a diameter of about 2.5 m looks down from space onto the night sky to detect near UV photons (330-400nm, both fluorescent and Cherenkov photons) emitted from the giant air showers produced by EECRs. The arrival direction map with more than five hundred events will tell us the origin of the EECRs and allow us to identify the nearest EECR sources with known astronomical objects. It will allow them to be examined in other astronomical channels. This is likely to lead to an understanding of the acceleration mechanisms, perhaps producing discoveries in astrophysics and/or fundamental physics. The comparison of the energy spectra among the spatially resolved individual sources will help to clarify the acceleration/emission mechanism, and also finally confirm the Greisen-Zatsepin-Kuz'min process for the validation of Lorentz invariance up to $\gamma \sim 10^{11}$. Neutral components (neutrinos and gamma rays) can also be detected as well, if their fluxes are high enough. The JEM-EUSO mission is planned to be launched by a H2B rocket about JFY 2015-2016 and transferred to ISS by H2 Transfer Vehicle (HTV). It will be attached to the Exposed Facility external experiment platform of "KIBO."

## 1. INTRODUCTION

The "Extreme Universe Space Observatory - EUSO" is the first space mission devoted to the exploration of the Universe through the detection of the extreme energy (E >100 ZeV) cosmic rays (EECRs) and neutrinos [1,2,3,4]; it looks downward from the International Space Station (ISS). It was first proposed as a free-flyer, but was selected by the European Space Agency (ESA) as a mission attached to the Columbus module of ISS. The phase-A study for the feasibility of that observatory (hereafter named ESA-EUSO) was successfully completed in July 2004. Nevertheless, because of financial problems in ESA and European countries, together with the logistic uncertainty caused by the Columbia accident, the start of the phase B had been pending.

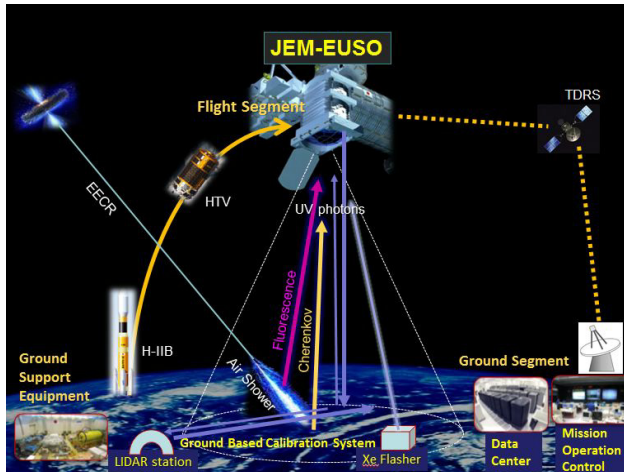

**FIGURE 1.** Principle of the JEM-EUSO telescope to detect Extreme Energy cosmic rays (EECRs).

In 2006, Japanese and U.S. teams redefined the mission as an observatory attached to "KIBO," the Japanese Experiment Module (JEM) of ISS. They renamed it JEM-EUSO and started with a renewed phase-A study.

JEM-EUSO is designed to achieve an exposure larger than 1 million km$^2$ sr year. This overwhelmingly high collecting power permits us to achieve our main scientific objective: astronomy and astrophysics through the particle channel to identify sources by arrival direction analysis and to measure the energy spectra from the individual sources, It will constrain acceleration or emission mechanisms, and also finally confirm the Greisen-Zatsepin-Kuz'min process [6] for the validation of Lorentz invariance up to $\gamma \sim 10^{11}$.

## 2. SCIENCE OBJECTIVES

Science objectives of the JEM-EUSO mission are divided into one main objective and five exploratory objectives. The main objective of JEM-EUSO is to initiate a new field of astronomy that uses the extreme energy particle channel ($5 \times 10^{19}$ eV $<$ E $< 10^{21}$ eV). JEM-EUSO has the critical exposure of 1 million km$^2 \cdot$ sr $\cdot$ year to observe all the sources at least once inside several hundred Mpc and makes possible the followings:

・ Identification of sources with the high statistics by arrival direction analysis
・ Measurement of the energy spectra from individual sources to constrain the acceleration or the emission mechanisms

We set five exploratory objectives:

・ Detection of extreme energy gamma rays
・ Detection of extreme energy neutrinos
・ Study of the Galactic magnetic field
・ Verification of the relativity and the quantum gravity effects at extreme energy
・ Global survey of nightglows, plasma discharges, and lightning

See [7,8] for the detailed discussions of scientific objectives.

## 3. INSTRUMENT

The JEM-EUSO instrument consists of the main telescope, an atmosphere monitoring system, and a calibration system. The main telescope of the JEM-EUSO mission is an extremely-fast ($\sim \mu$ s) and highly-pixelized ($\sim 3 \times 10^5$ pixels) digital camera with a large diameter (about 2.5m) and a wide-FoV($\pm 30°$). It works in near-UV wavelength (330-400 nm) with single-photon-counting mode. The telescope consists of four parts: the optics, the focal surface detector and electronics, and the structure. The optics focuses the incident UV photons onto the focal surface with an angular resolution of 0.1 degree. The focal surface detector converts the incident photons to photoelectrons and then to electric pulses.

Two curved double sided Fresnel lenses with 2.65m external diameter, an intermediate curved precision Fresnel lens, and a pupil constitute "baseline" optics of the JEM-EUSO telescope. The Fresnel lenses provide a large-aperture, and a wide FoV as well as a low mass and a high UV light transmittance.

Combination of three Fresnel lenses realized a full angle FoV of 60° and an angular resolution of 0.1°. This resolution corresponds approximately to (0.75 - 0.87) km on the earth's surface, depending on the location inside the FoV in nadir pointing mode. The material of the lens is UV transmitting PMMA which has a high UV transparency in the wavelength from 330nm to 400nm. Prototype sample is shown in Fig. 2. A precision Fresnel optics adopting a diffractive optics technology is used to suppress the color aberration. Details of the optics are described in [9].

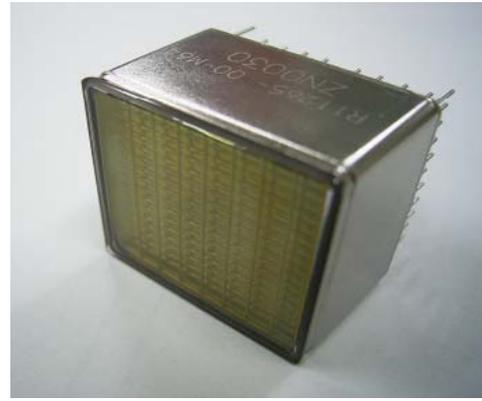

**FIGURE 3:** A baseline MAPMT, M64

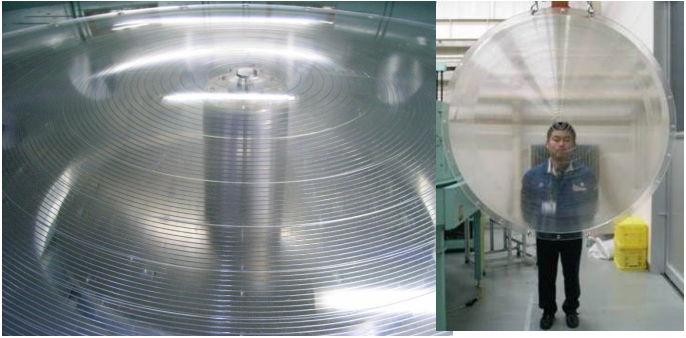

**FIGURE 2:** The picture of the rear lens: close up (left) and whole lens (right)[10]

The focal surface (FS) of JEM-EUSO is a sphere with about 2.7 m curvature radius, and it is covered with about 5,000 multi-anode photomultiplier tubes. The FS detector consists of Photo-Detector Modules (PDMs), each of which consists of nine Elementary Cells (ECs). The EC contains four units of the MAPMTs. About 137 PDMs are arranged in FS. Details of the focal surface detector are described in [11,12]. We will use weakly focused MAPMTs, M64 with 8x8=64 pixels (Figure 3).

The FS electronics system records the signals of UV photons generated by EECRs in time. The system is required to keep a high trigger efficiency with a flexible trigger algorithm as well as a reasonable linearity over $10^{19}$-$10^{21}$ eV range. Power consumption per channel is required to be less than 2.5mW to manage 3x $10^5$ signal channels in an available power budget (1kW).

The FS electronics are configured in three levels: Front-end electronics at an EC level, PDM electronics for nine EC units, and FS electronics to control 137 units of PDM electronics. Anode signals of the MAPMT are counted and recorded in ring memories for each GTU (Gate Time Unit~2.5 $\mu$ s) to wait for a trigger assertion, then, the data are read and sent to control boards. JEM-EUSO uses hierarchical trigger scheme to reduce data rate of ~10GB/s/FS down to 297 kbps for sending data from ISS to ground operation center.

A PDM operates in the following four trigger modes: a) Normal mode with a GTU of 2.5 $\mu$ s for routine data taking of EAS, b) Slow mode with a programmable GTU up to a few ms, for the study of meteorites and other atmospheric luminous phenomena, c) Detector calibration mode with a GTU value suitable for the calibration runs, and d) Lidar mode with a GTU of 200ns. Details of the FS electronics system are described in [13,14,15].

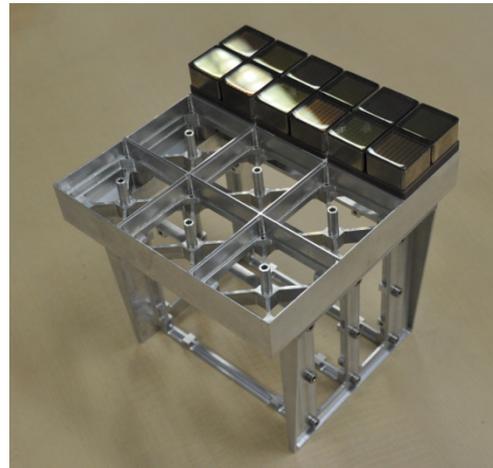

**FIGURE 4:** Bread board model of the PDM structure. Twelve PMTs are attached on the structure, here. Nine ECs and six electronics board will be accommodated on the structure.

Atmosphere Monitoring System (AMS) monitors the earth's atmosphere continuously inside the FoV of the JEM-EUSO telescope. Intensity of the fluorescent and Cherenkov light emitted from EAS observed in JEM-EUSO depends on the transparency of the atmosphere, the cloud coverage, the height of cloud top, albedo of the ground, and other atmospheric and ground conditions.

The AMS uses IR camera, Lidar, and the slow data of the main telescope to measure the cloud-top height with accuracy better than 500 m. An IR camera observes the infrared fluxes of spectral bands in 11-13 $\mu$ m to determine the temperature of the cloud top. The cloud height is calculated from the temperature with the conversion table, which is calibrated by the ranging data of the Lidar system. The Laser of the JEM-EUSO Lidar releases the short pulse (less than 10ns, 20mJ/pulse) of the UV photons with 355nm in the frequency of 50Hz. The returned pulses are observed by the main telescope in a higher time resolution (200ns) of the Lidar mode of the PDM. The slow data of the main telescope can also be used to determine the cloud top height by trigonometric parallax. Details of AMS are described in [16].

The calibration system measures the efficiencies of the optics, the focal surface detector, and the data acquisition electronics. The pre-flight calibration of the detector will be done by measuring detection efficiency, uniformity, gain etc. with UV LED's for several kinds of wavelength. On the other hand, few diffuse light sources of LEDs with different wavelengths in the near UV, are placed at the support of the rear lens before FS and illuminate FS to measure efficiencies and gain of FS detector on-board. Similar light sources are placed on FS to measure efficiencies of the lenses. Reflected light at the inner surface of the lid is observed with FS. In this way, the gain and the detection efficiency of the detector will be calibrated on board.

To accommodate JEM-EUSO into a volume of the HTV transfer vehicle, a contractible/extensible structure is adopted. The structure is stowed at launch by H2B rocket and it is extended at ISS.

## 4. EXPECTED PERFORMANCE

The performance of the instrument is evaluated by the end-to-end simulations. The threshold energy of JEM-EUSO in nadir mode (50% trigger efficiency) is $5 \times 10^{19}$ eV, and the trigger efficiency at $10^{20}$ eV is 86% for the typical background rate. The threshold energy for showers observed in a FoV of 15 degrees can be lowered down to $3.7 \times 10^{19}$ eV because of a better efficiency of the optics in the center of the FoV, and due to the smaller distance of the EAS axis to the detector. The threshold energies rise by the tilt mode observation due to longer distance to EAS axis. However, the increase of the acceptance by the tilt mode observation has an important advantage over it. Details of the expected performance are described in [17].

## 5. CONCLUSION

JEM-EUSO is the science mission looking downward from the ISS to explore the extremes in the Universe and fundamental physics through the detection of the extreme energy E > $10^{20}$ eV) cosmic rays. It is the first instrument that has a full-sky coverage and achieves an exposure more than one million km$^2$・sr・year, the critical value of the exposure to start "astronomy and astrophysics through particle channel." The JEM-EUSO mission is planned to be launched by a H2B rocket about 2015-2016 and transferred to ISS by H2 transfer vehicle (HTV), and attached to the external experiment platform of "KIBO."